\documentclass[sigconf,screen]{acmart}

\usepackage{xr}
\usepackage{multirow}
\usepackage{caption}
\usepackage{subcaption}
\usepackage{graphicx}
\usepackage{cleveref}
\usepackage{xcolor}
\usepackage{soul}
\usepackage{color, colortbl}
\definecolor{Gray}{gray}{0.7}
\usepackage{amsmath}
\usepackage{mathtools}

\newlength\myboxwidth
\AtBeginDocument{%
  \providecommand\BibTeX{{%
    \normalfont B\kern-0.5em{\scshape i\kern-0.25em b}\kern-0.8em\TeX}}}

\copyrightyear{2024}
\acmYear{2024}
\setcopyright{rightsretained}
\acmConference[CHI EA '24]{Extended Abstracts of the CHI Conference on Human Factors in Computing Systems}{May 11--16, 2024}{Honolulu, HI, USA}
\acmBooktitle{Extended Abstracts of the CHI Conference on Human Factors in Computing Systems (CHI EA '24), May 11--16, 2024, Honolulu, HI, USA}
\acmDOI{10.1145/3613905.3650981}
\acmISBN{979-8-4007-0331-7/24/05}

\begin{document}

\title{An Interactive Tool for Simulating Mid-Air Ultrasound Tactons on the Skin}



\author{Chungman Lim}
\email{chungman.lim@gm.gist.ac.kr}
\orcid{0000-0002-7857-3322}
\affiliation{%
  \institution{Gwangju Institute of\\ Science and Technology}
  \country{Republic of Korea}}

\author{Hasti Seifi}
\email{hasti.seifi@asu.edu}
\orcid{0000-0001-6437-0463}
\affiliation{%
  \institution{Arizona State University}
  \city{Tempe}
  \country{United States}}

\author{Gunhyuk Park}
\email{maharaga@gist.ac.kr}
\orcid{0000-0003-2677-5907}
\affiliation{%
  \institution{Gwangju Institute of\\ Science and Technology}
  \country{Republic of Korea}
}



\begin{abstract}

Mid-air ultrasound haptic technology offers a myriad of temporal and spatial parameters for contactless haptic design. 
Yet, predicting how these parameters interact to render an ultrasound signal is difficult before testing them on a mid-air ultrasound haptic device.  
Thus, haptic designers often use a trial-and-error process with different parameter combinations to obtain desired tactile patterns (i.e., Tactons) for user applications.
We propose an interactive tool with five temporal and three spatiotemporal design parameters that can simulate the temporal and spectral properties of stimulation at specific skin points.
As a preliminary verification, we measured vibrations induced from the ultrasound Tactons varying on one temporal and two spatiotemporal parameters.
The measurements and simulation showed similar results for three different ultrasound rendering techniques, suggesting the efficacy of the simulation tool.  
We present key insights from the simulation and discuss future directions for enhancing the capabilities of simulations.

\end{abstract}




\begin{teaserfigure}
  \centering
  \includegraphics[width=\textwidth]{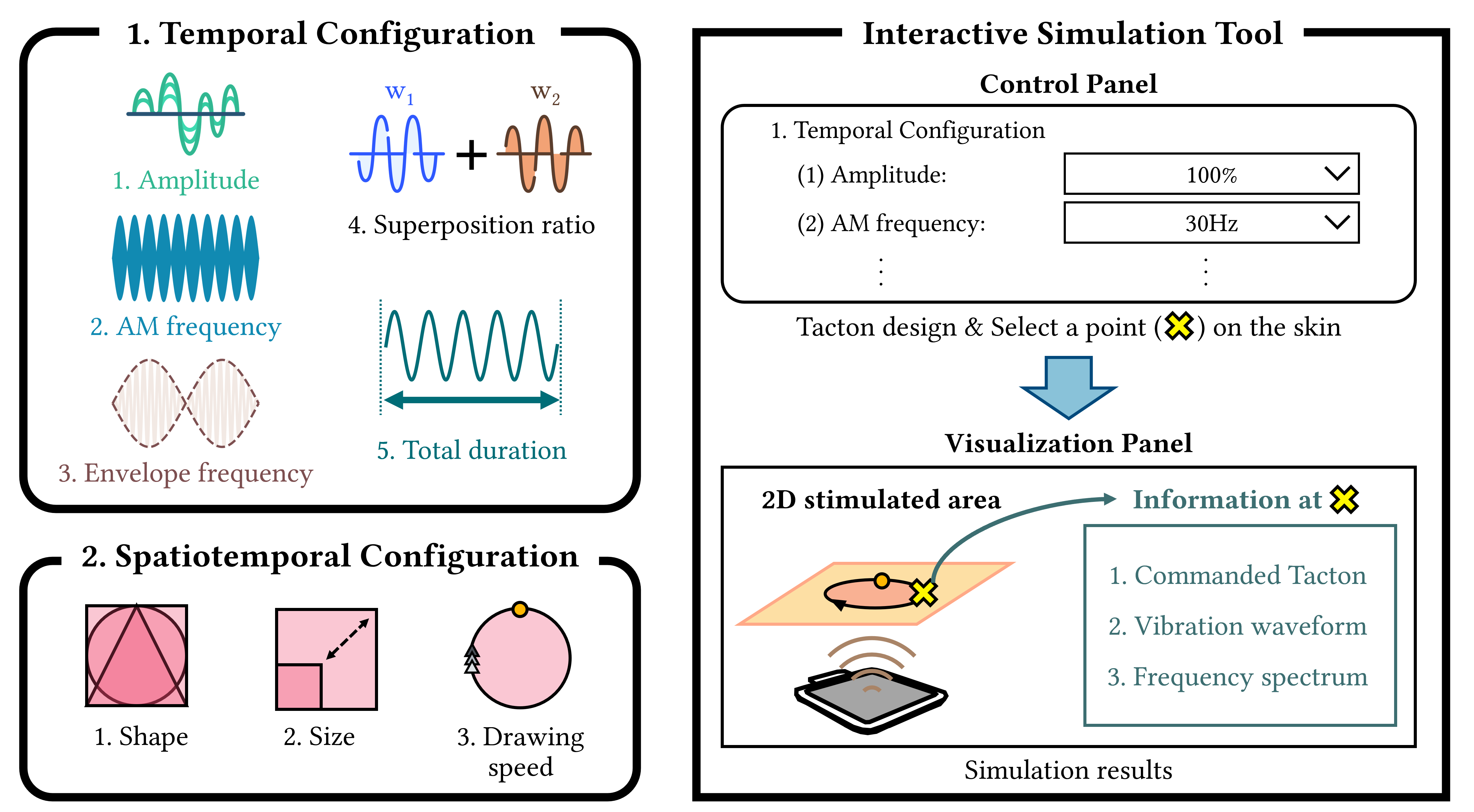}
  \caption{ 
  Overview of our interactive simulation tool for testing design parameters in mid-air ultrasound technology.
  Haptic designers can control five temporal and three spatiotemporal parameters to design a mid-air ultrasound Tacton.  
  They can also select a point on the 2D plane to visualize the temporal waveform and frequency spectrum of the Tacton at that point on the skin.
  }
  \label{fig:overview}
  \Description{}
\end{teaserfigure}

\sloppy
\maketitle

\section{Introduction}

Mid-air ultrasound technologies create haptic feedback on the user's skin without physical contact with a device.
This technology focuses acoustic waves into one or multiple focal points using a phased array of ultrasonic transducers, and modulates or moves these focal points to create a sense of touch~\cite{rakkolainen2020survey}.
Designers are exploring the parameter space of the mid-air ultrasound patterns (i.e., tactile icons or Tactons) to deliver information or emotion to users in various applications, including touchless public displays~\cite{limerick2019user, vi2017not}, automotive user interfaces~\cite{harrington2018exploring, brown2022augmenting}, medical training simulations~\cite{hung2013ultrapulse, hung2014using}, and virtual reality environments~\cite{howard2022ultrasound, hwang2017airpiano, mulot2023ultrasound, villa2022extended}.

Several rendering techniques and parameters exist for creating mid-air ultrasound Tactons.
The common approaches include using either temporal parameters (e.g., amplitude-modulated frequency)~\cite{hoshi2009non}, spatiotemporal parameters (e.g., trajectory or drawing speed of a focal point)~\cite{frier2018using, takahashi2019tactile, mulot2023improving}, or a combination of both~\cite{hajas2020mid, rutten2020discriminating, dalsgaard2022user}.
These rendering techniques accompany complex physical effects on the skin. 
For example, amplitude modulation (AM) focuses acoustic pressures on a static focal point, vibrating the local distribution of skin and activating a group of mechanoreceptors.
Spatiotemporal modulation (STM) moves a focal point along a trajectory, thus vibrating the skin at a drawing frequency on the trajectory points.
Furthermore, the combination of AM and STM techniques offers both AM frequency and drawing frequency, making it challenging for haptic designers to predict the rendered frequency on the user's palm.

The combinations of parameters from the above techniques usually yield in vibrations that are complex to predict, so the designers resort to repeat trial-and-errors to find a Tacton set of their interests.
Thus, the physical simulations of the mid-air ultrasound Tacton can help understand its perception and reduce the Tacton design cost.
Prior literature proposes physical simulations for the ultrasound vibrations and their measurement data, for example, AM or STM of a focal point~\cite{carter2013ultrahaptics, chilles2019laser} and gap detection thresholds between two static focal points~\cite{carter2013ultrahaptics, howard2023gap}.
Yet, as the existing simulations typically simulate one parameter or one rendering technique, more research is needed on interactive simulation tools for testing the interaction of temporal and spatiotemporal parameters in complex ultrasound Tactons and the combination of rendering techniques.

To fill this gap, we developed a Python-based interactive simulation tool for skin vibrations induced by ultrasound Tactons rendered by modulating a single focal point.
Our simulation facilitates the design of Tactons with five temporal parameters: amplitude, AM frequency, envelope frequency, superposition ratio, and total duration; and three spatiotemporal parameters: the shape, size, and drawing speed of a focal point's trajectory. 
These parameters are commonly used by designers for creating Tactons.
Designers can manipulate these eight parameters in the control panel and view the vibration waveform and frequency spectrum at any point on the skin in the visualization panel.
Thus, the simulation tool enables designers to efficiently explore the physical effects of multiple parameters before rendering the Tactons on the device.

As an initial verification of the simulation tool, we designed 15 mid-air ultrasound Tactons varying in AM frequency, size, and drawing speed and tested AM rendering, STM rendering, and a combination of AM and STM rendering techniques.
We selected the three parameters in our preliminary measurements because they mainly affect the spectral peaks of the induced vibrations at a skin point.  
We used our simulation tool to predict vibrations at five points on the skin.
Then, we employed a STRATOS Explore ultrasound haptic device to render the Tactons and measured vibrations induced at these points with a laser vibrometer. 
Our preliminary measurements showed high correspondence with our simulation results, revealing similar temporal waveforms and spectral harmonics to the simulated predictions.
We discuss directions for future research to improve the measurement methodology and to build a high-utility simulation.
Our contributions include:
\begin{itemize}
    \item An interactive tool that simulates the vibrations induced on the skin from physical interactions of five temporal and three spatiotemporal parameters in mid-air ultrasound Tactons.
    \item Preliminary measurements of vibrations induced by 15 mid-air ultrasound Tactons, suggesting the validity of the simulation tool for AM and STM rendering.
\end{itemize}

\section{Method and implementation}

We developed a computational model to simulate mid-air ultrasound Tactons that use a single focal point.
This model predicts the vibration waveforms and frequency spectra at any points on the skin area.

\subsection{Assumptions for Simulation}
\label{sec:assumptions}
We made several assumptions about mid-air ultrasound stimulation in order to lower the computational complexity of the simulation tool. 
In the literature, focused mid-air ultrasound at a single focal point creates a central oval-shaped vibration area with maximum amplitude, followed by four smaller vibration areas located at four directions apart 15\,$cm$ from the center with about less than 30 percent of the center amplitude (side lobes) if the ultrasound device located at the 20\,$cm$ distance~\cite{carter2013ultrahaptics, wilson2014perception}.
The amplitudes of side lobes are nearly below the detection threshold at the maximum stimulation~\cite{howard2019investigating}, while the side lobes vary on the distance between the device and the focal point due to changes in the angles of the ultrasound transducers.
Therefore, our model assumes the ultrasound stimulation occurs at a target single circular point on the skin.
In addition, we assumed that the skin of the user's palm is a 2D plane, and the propagation of waves along the skin does not occur, considering the computational complexity of the model.
We discuss the implications of these assumptions in Section~\ref{sec:discussion}.

\subsection{Parameter Space of Mid-Air Ultrasound Tactons in Simulation}

\begin{figure*}[t]
  \centering

    \begin{subfigure}[c]{0.3\textwidth}
      \includegraphics[width=\linewidth]{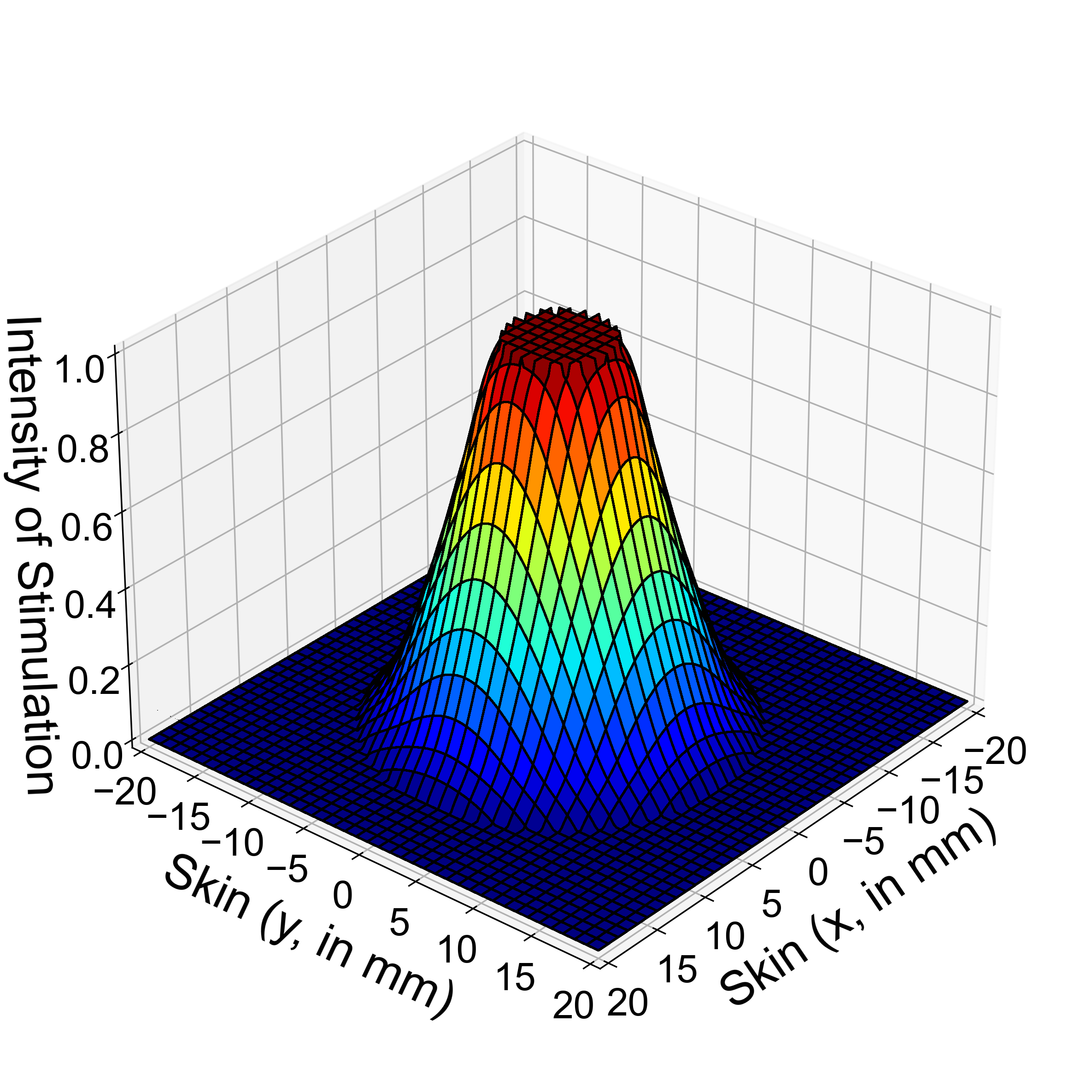}
    \end{subfigure}
    \quad
    \quad
    \quad
    \begin{subfigure}[c]{0.27\textwidth}
      \includegraphics[width=\linewidth]{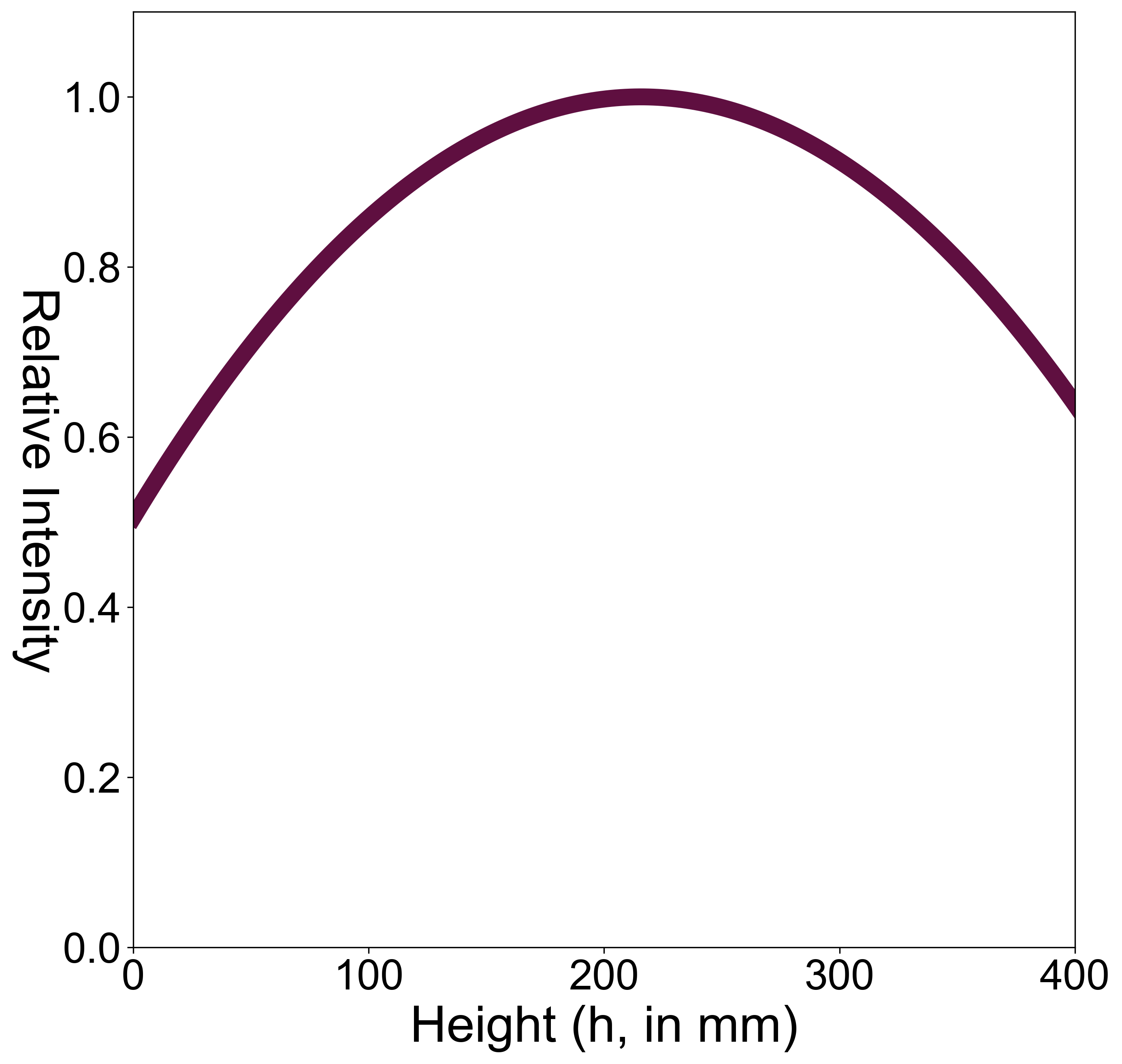}
    \end{subfigure}

    \begin{subfigure}[c]{0.3\textwidth}
      \caption{Intensity by a static focal point}
      \label{fig:Stimulation}
    \end{subfigure}
    \quad
    \quad
    \quad
    \begin{subfigure}[c]{0.27\textwidth}
      \caption{Relative intensity by height from a device}
      \label{fig:Height}
    \end{subfigure}

    \begin{subfigure}[c]{0.32\textwidth}
    \includegraphics[width=\linewidth]{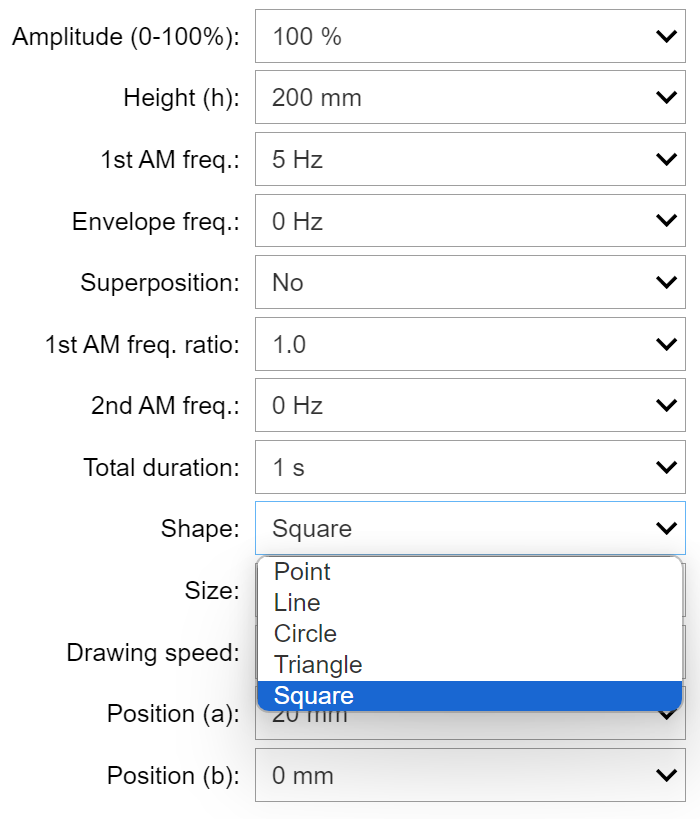}
    \end{subfigure}
    \hfill
    \begin{subfigure}[c]{0.67\textwidth}
    \includegraphics[width=\linewidth]{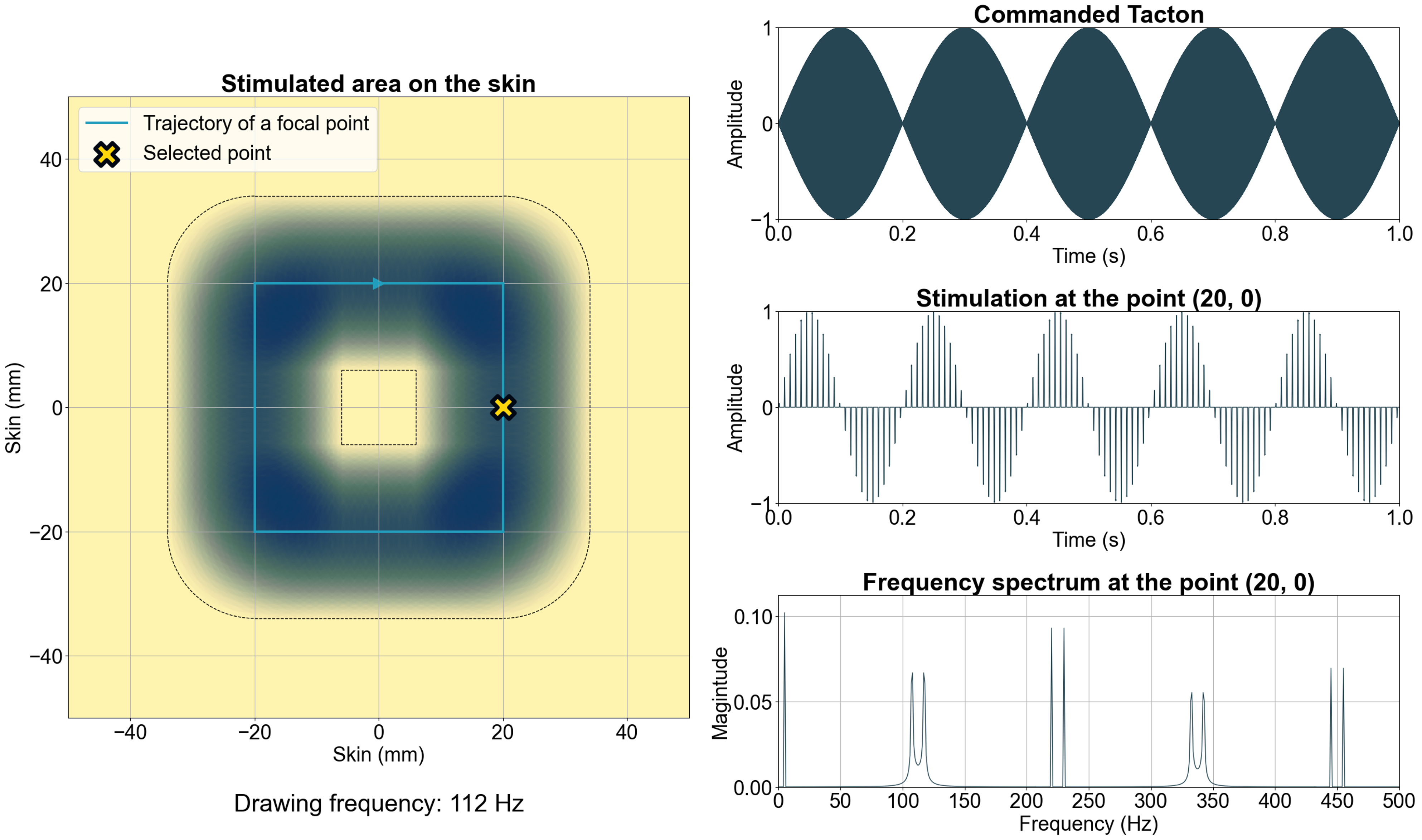}
    \end{subfigure}

    \begin{subfigure}[c]{0.32\textwidth}
      \caption{Control panel}
      \label{fig:ControlPanel}
    \end{subfigure}
    \hfill
    \begin{subfigure}[c]{0.67\textwidth}
      \caption{Visualization panel}
      \label{fig:VisualizationPanel}
    \end{subfigure}

  \caption{
  Plots for the intensity of stimulation by a single focal point at the height h and the interactive simulation tool:
  (a) The intensity of stimulation decreases with distance from the focal point~\cite{carter2013ultrahaptics}.
  (b) The relative intensity determined by the distance between a focal point and the mid-air ultrasound device.
  (c) The control panel in our interactive tool for manipulating parameters in temporal and spatiotemporal configurations. The dropdowns allow users to select design parameters and a position ($\vec{a} = (a,~b)$) on the skin.
  (d) The visualization panel showing the 2D stimulated skin area (Left) and vibration waveform (Right) at a specific point on the skin. The sky-colored line represents the trajectory of a focal point, the black dotted line represents the borderline influenced by the stimulation, and the ``X'' symbol represents the point selected to see the effects of the stimulation by the Tacton.
  The three plots (Right) display the temporal plot of the Tacton, and the temporal and spectral plots of the stimulation at the selected point $\vec{a}$ by the user .
  }

  \vspace*{-8pt}
\end{figure*}

\begin{figure*}[t]

    \centering
    \begin{subfigure}[c]{0.4\textwidth}
    \includegraphics[width=\linewidth]{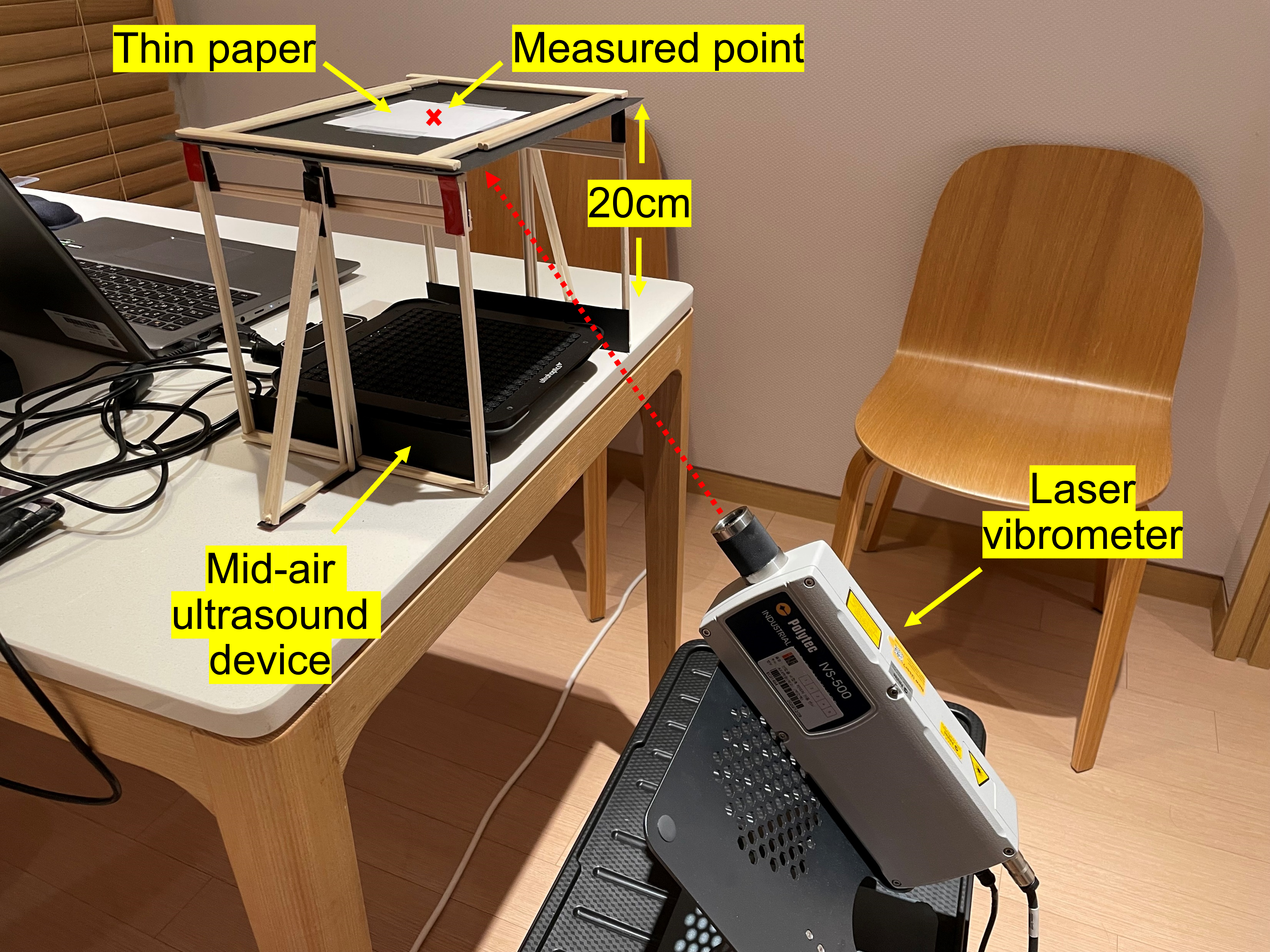}
    \end{subfigure}
    \quad
    \quad
    \quad
    \begin{subfigure}[c]{0.3\textwidth}
    \includegraphics[width=\linewidth]{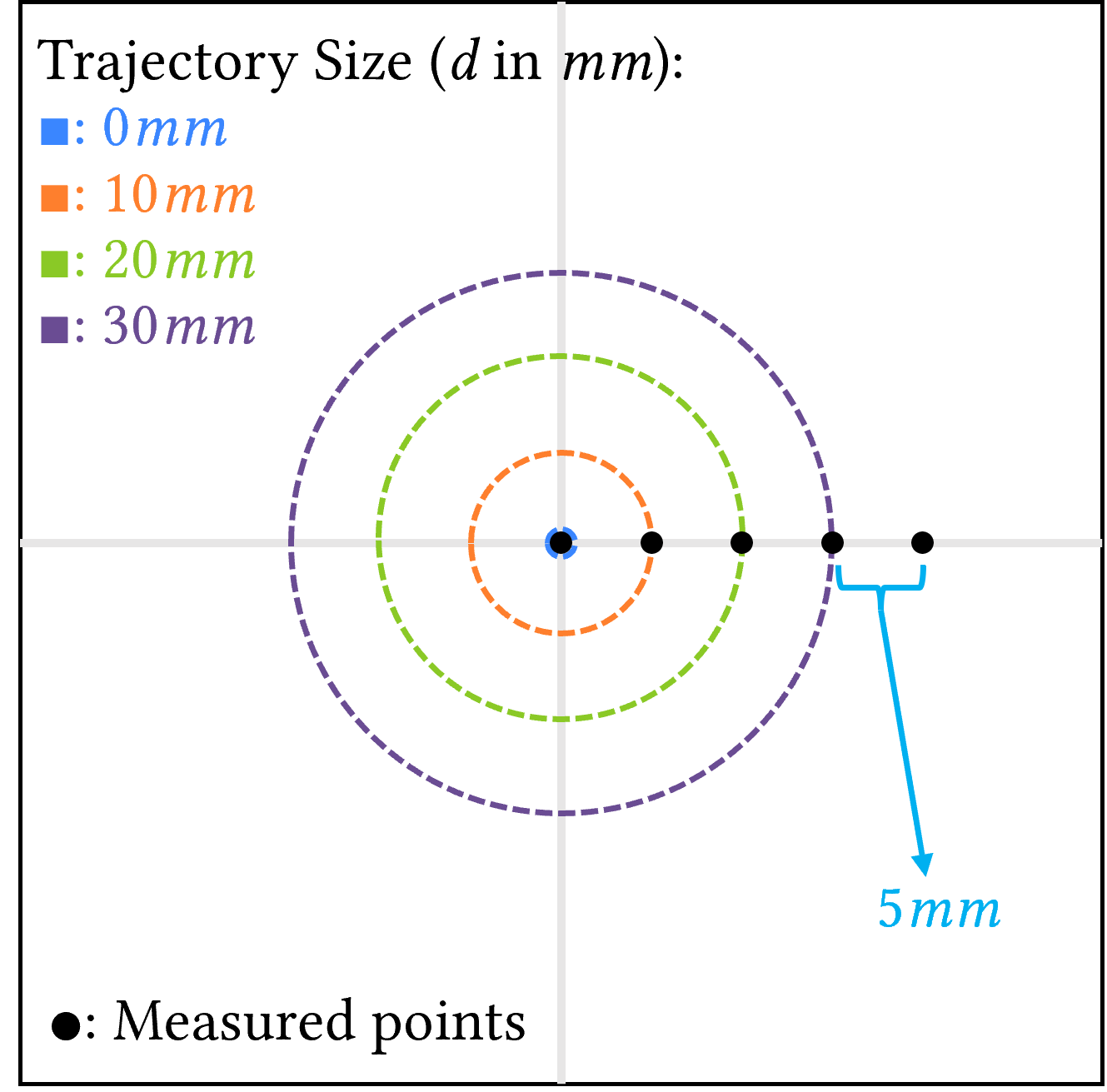}
    \end{subfigure}

    \begin{subfigure}[c]{0.4\textwidth}
    \caption{}
    \label{fig:setup}
    \end{subfigure}
    \quad
    \quad
    \quad
    \begin{subfigure}[c]{0.3\textwidth}
    \caption{}
    \label{fig:Mpoint}
    \end{subfigure}
    
    \caption{
    Our measurement setup. (a) We used a laser vibrometer to measure vibrations on paper induced by mid-air ultrasound Tactons. (b) In all measurements, we measured the vibrations at the same 5 points, spaced at 5 mm intervals.
    }

    \vspace*{-10pt}
\end{figure*}

Our model provides two configuration spaces for rendering mid-air ultrasound Tactons (Figure~\ref{fig:overview}): (1) The temporal configuration includes five parameters of \textit{amplitude}, \textit{AM frequency}, \textit{envelope frequency}, \textit{superposition ratio}, and \textit{total duration}; and (2) the spatiotemporal configuration includes three parameters of \textit{shape}, \textit{size}, \textit{drawing speed} of a focal point trajectory.
We selected these parameters based on the most frequently used parameters in Tacton design literature~\cite{dalsgaard2022user, hajas2020mid, obrist2015emotions, obrist2013talking, hwang2017perceptual, seifi2015vibviz, park2011perceptual}.
Thus, users can observe the vibration waveform at any point on the skin by controlling these eight parameters of mid-air ultrasound Tactons.

\textbf{1. Temporal configuration:}
We defined the temporal configuration of mid-air ultrasound Tactons using the mathematical equation $p(t)$, which controls five parameters:

\begin{equation}
\fontsize{7}{9}\selectfont
    p (t) =
    \begin{dcases}
        A \cdot U(t) \{w_{AM_1} M_1 (t) + w_{AM_2} M_2 (t)\} E(t), &\!\!\! \text{if \textit{superposition ratio} is used.} \\
        A \cdot U(t) M(t) E(t), & \text{otherwise.}
    \end{dcases}
    \label{equ:temporalConfiguration_1}
\end{equation}

where $U(t)$ represents the continuous ultrasound at 40\,kHz (typical frequency from commercial mid-air ultrasound haptic devices), $M(t)$ is an AM sinusoid of $sin(2\pi f_{AM} t)$ with AM frequency $f_{AM}$, $E(t)$ is an envelope sinusoid of $sin(2\pi f_e t)$ with envelope frequency $f_e$, and $w_{AM_{1}}$ and $w_{AM_{2}}$ is the weights of two AM sinusoids (i.e., $M_1 (t)$ and $M_2 (t)$).

\looseness-1 \textit{Amplitude} ($A$) corresponds to the peak acoustic pressure at the focal point, ranging between 0\% and $\pm$~100\%, as provided by a mid-air ultrasound device. 
While $A$ represents a commanded amplitude on the device, its relative intensity varies with height, which is the vertical distance between the focal point and the mid-air ultrasound device.
The amplitude is known to have an inverted U-shaped relationship with height, reaching a maximum at 200~$mm$~\cite{raza2019perceptually} above the device.  
We applied the above findings to calculate relative intensity of $A$ as in Figure~\ref{fig:Height}.
\textit{AM frequency} ($f_{AM}$) represents a temporal frequency that modulates $U(t)$~\cite{obrist2013talking, obrist2015emotions}. 
Here, $f_{AM}$ = 0\,Hz indicates no AM rendering (i.e., $M(t) = 1$), suggesting that STM rendering is necessary to create tactile sensations.
\textit{Envelope frequency} ($f_{e}$) refers to a frequency modulating $M(t)$~\cite{park2011perceptual} and $f_{e}$ = 0\,Hz denotes a constant envelope (i.e., $E(t) = 1$).
\textit{Superposition ratio} ($w_{AM_{1}}$:$w_{AM_{2}}$) is the mixing ratio of two sinusoidal signals for creating a superimposed signal.
The simulation tool currently provides five ratios: 1:0, 0.75:0.25, 0.5:0.5, 0.25:0.75, 0:1~\cite{yoo2022perceived, hwang2017perceptual, lim2023can}.
\textit{Total duration} ($t_{d}$) is the maximum $t$ of the vibrations.
While the computational model can handle any duration, we set 10 seconds as the maximum in the current tool, in line with the Tacton durations common in user applications~\cite{seifi2015vibviz}.

\textbf{2. Spatiotemporal configuration:}
We defined the spatiotemporal configuration as the spatial properties of the temporal formula $p(t)$, consisting of three parameters, \textit{shape}, \textit{size}, and \textit{drawing speed} of a focal point trajectory in a 2D plane above the mid-air ultrasound device (Figure~\ref{fig:overview}).
In this configuration, \textit{shape} represents the trajectory of a focal point in the 2D plane.
Based on the literature~\cite{rutten2019invisible, hajas2020mid}, we provide five shapes: point, horizontal line, circle, regular triangle, and square.
\textit{Size} ($d$, in $mm$) refers to the length of the line, diameter of the circle, or the side length for the regular triangle and square.
We set the maximum \textit{size} as 60\,$mm$, considering the typical size of a user's palm~\cite{shen2023multi, hajas2020mid}.
\textit{Drawing speed} ($v$, in $m/s$) denotes the velocity of a focal point's movement.
Combinations of these three parameters derive a drawing frequency ($f_{d}$), defined as the number of completions or revolutions of a trajectory per second~\cite{freeman2021perception, wojna2023exploration, rutten2020discriminating}, creating a spatiotemporal tactile sensation (i.e., STM)~\cite{frier2018using}.
The three parameters of \textit{shape}, \textit{size}, and \textit{drawing speed} determine the 2D position of a focal point on the moving trajectory at $t$, represented as $\vec{x}(t)$ = ($x(t)$,~$y(t)$).
For example, when \textit{shape} is the point, $\vec{x}(t)$ is a constant at (0,~0).
When \textit{shape} is the circle, $\vec{x}(t)$ can be expressed as ($\frac{d}{2} \cos(2 \pi f_{d} t),~\frac{d}{2} \sin(2 \pi f_{d} t)$).
With the trajectory $\vec{x}(t)$ from a total of the three parameters, we denoted $p(t)$ at the focal point as $p(\vec{x}(t),t)$.

\begin{figure*}[t]
  \centering

  \begin{subfigure}[t]{0.3\textwidth}
    \includegraphics[width=\linewidth]{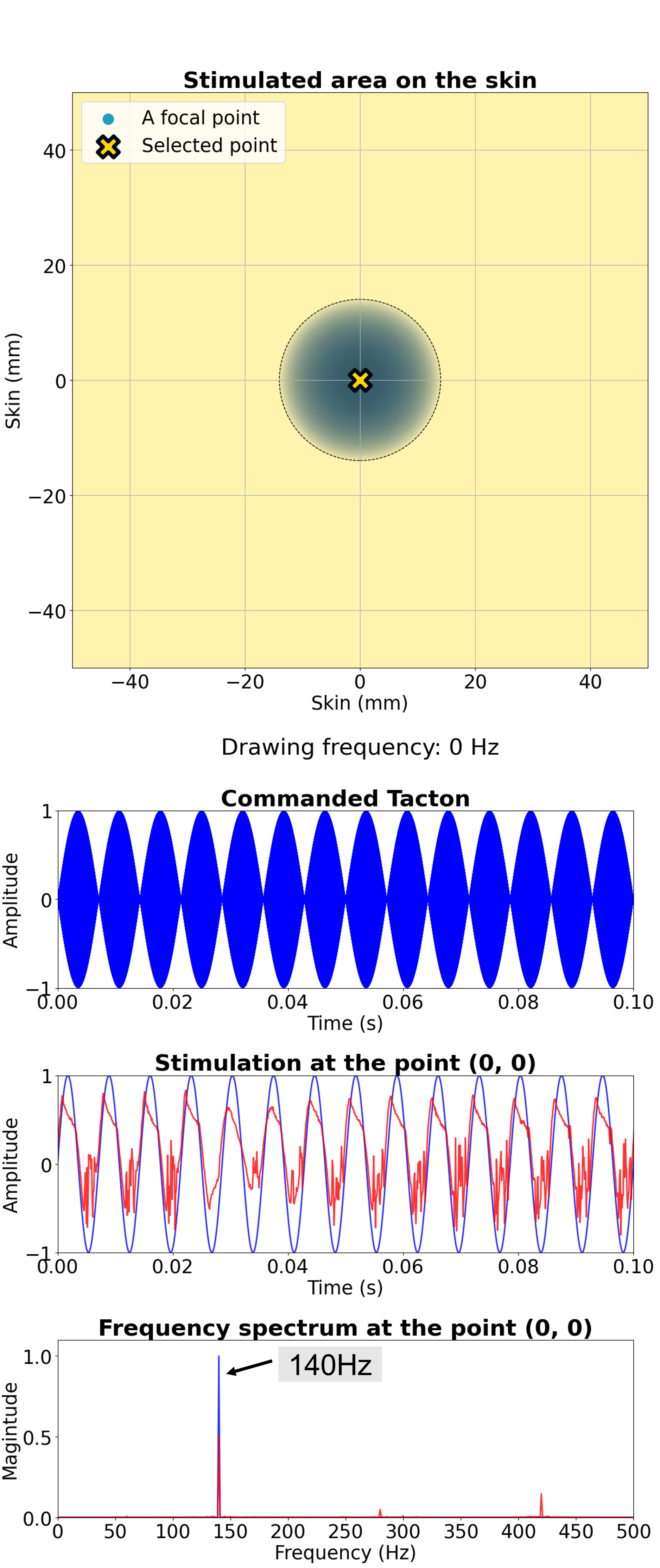}
    \caption{$PureAM$ \\ ($f_{AM}$ = 140\,Hz)}
    \label{fig:Comparison_AM}
  \end{subfigure}%
  \hfill
  \begin{subfigure}[t]{0.3\textwidth}
    \includegraphics[width=\linewidth]{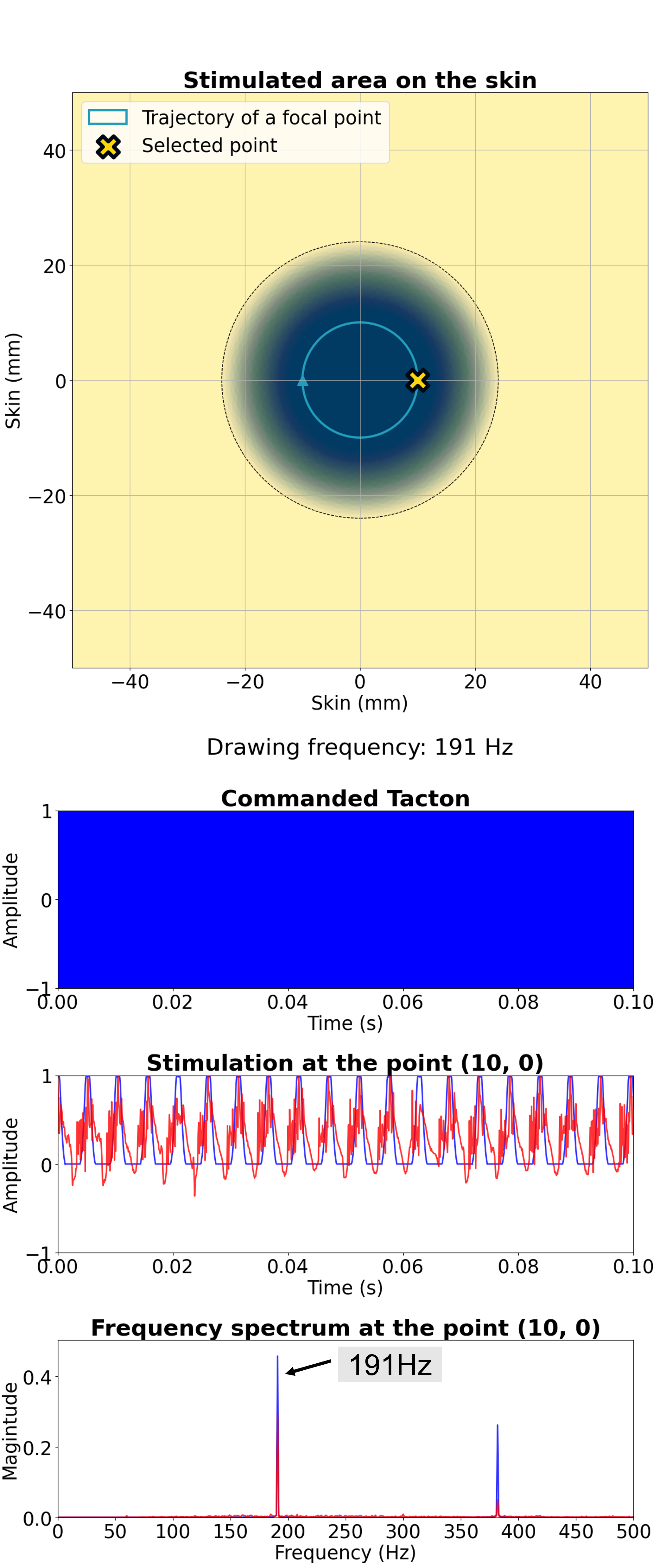}
    \caption{$PureSTM$ ($d$~=~20\,$mm$, $v$~=~12\,$m/s$, $f_{d}$~=~191\,Hz)}
    \label{fig:Comparison_STM}
  \end{subfigure}
    \hfill
  \begin{subfigure}[t]{0.3\textwidth}
    \includegraphics[width=\linewidth]{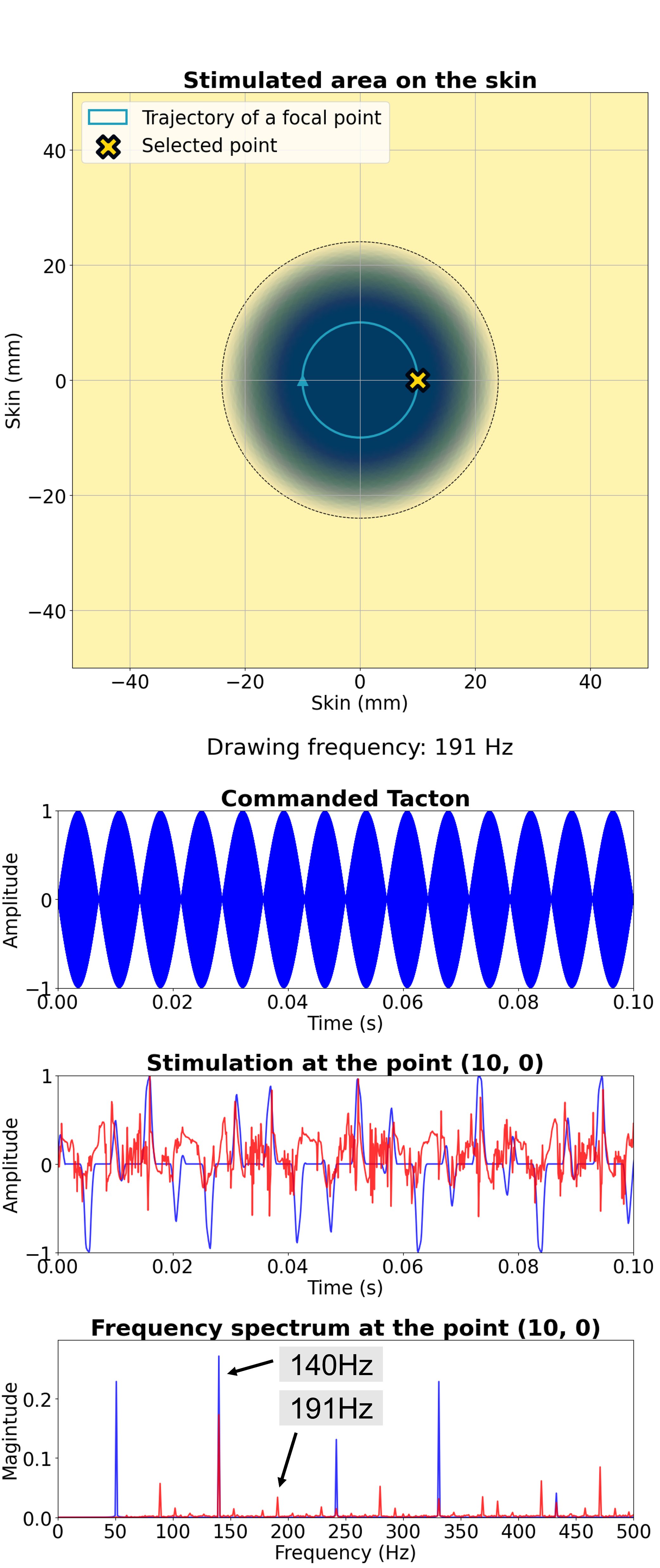}
    \caption{$AM{+}STM$ ($f_{AM}$~=~140\,Hz, $d$~=~20\,$mm$, $v$~=~12\,$m/s$, $f_{d}$~=~191\,Hz)}
    \label{fig:Comparison_AM_STM}
  \end{subfigure}

  \caption{Three exemplar comparisons for pure AM rendering, pure STM rendering, and combination of AM and STM rendering, between simulations (blue) and measurements (red).}
  \label{fig:Comparison}
    \vspace*{15pt}
\end{figure*}

\vspace*{-4pt}
\subsection{Model Architecture}
At a selected point $\vec{a} = (a,b)$ on the 2D plane above the device, we can define the Euclidean distance between $\vec{x}(t)$ (location of the focal point) and $\vec{a}$ as $D(\vec{x}(t),\vec{a})$.
The distance determines the intensity of stimulation at $\vec{a}$, which we defined as $S(D(\vec{x}(t),\vec{a}))$ (Figure~\ref{fig:Stimulation}).
Thus, we defined the final intensity of stimulation as $A(\vec{x}(t),\vec{a}) = A \cdot S(D(\vec{x}(t),\vec{a}))$ and expressed $p(\vec{x}(t),t)$ at $\vec{a}$ as:

\begin{equation}
\fontsize{5.9}{8.9}\selectfont
    p(\vec{x}(t),\vec{a},t) {=} 
    \begin{dcases}\!
        A(\vec{x}(t),\vec{a}) U(t) \{w_{AM_1} M_1 (t) {+} w_{AM_2} M_2 (t)\} E(t), &\!\!\!\! \text{if \textit{superposition ratio} is used.} \\
        A(\vec{x}(t),\vec{a}) U(t) M(t) E(t), &\text{otherwise.} 
    \end{dcases}
    \label{equ:spatiotemporalConfiguration_2}
\end{equation}

\looseness-1  With this implementation, our model facilitates tests for the temporal and spatiotemporal configurations, comprising five and three parameters, respectively (Figure~\ref{fig:ControlPanel}).
In the interactive simulation tool, we provide the stimulated areas on the skin by $p(\vec{x}(t),t)$ (Figure~\ref{fig:VisualizationPanel} Left).
Our tool also presents the temporal plot of the commanded Tacton (i.e., Equation~\ref{equ:temporalConfiguration_1}) considering $U(t)$ as 40\,$kHz$ sinusoid.
For the temporal plot for the signal at $\vec{a}$ (i.e., Equation~\ref{equ:spatiotemporalConfiguration_2}), we substituted 1 to $U(t)$ because the ultrasound stimulation acts as a constant pressure~\cite{frier2018using, shen2023multi}.
Then we applied Fourier transform to the temporal waveform to estimate the frequency spectrum (Figure~\ref{fig:VisualizationPanel} Right).

\section{Preliminary Measurement}

For preliminary verification for the capability of our simulation tool, we designed and measured 15 mid-air ultrasound Tactons considering three rendering scenario: (1) Pure AM rendering ($PureAM$), (2) pure STM rendering ($PureSTM$), and (3) a combination of AM and STM ($AM{+}STM$).

\subsection{Methods}

We selected ultrasound Tactons for the measurements, focusing on \textit{AM frequency} ($f_{AM}$) in the temporal configuration and \textit{size} ($d$) and \textit{drawing speed} ($v$) in the spatiotemporal configuration, as these three parameters mainly determine the AM frequency and the drawing frequency of the ultrasound Tactons which affects vibration spectrum.
We did not use the other temporal parameters and we kept \textit{total duration} at 1 second.
In addition, we maintained \textit{shape} as the circle and \textit{height} at 200\,$mm$.
We used four $f_{AM}$ values: 0, 80, 140, and 210\,Hz.
We selected $f_{AM}$ = 0\,Hz to test pure STM rendering ($PureSTM$) and $f_{AM}$ > 60\,Hz, as the power lines in our country introduces a constant 60\,Hz measurement noise.
We also used four combinations of size and drawing speed as ($d$~in~$mm$, $v$~in~$m/s$): (0,~0), (10,~6), (20,~12), (30,~18).
We chose (0\,$mm$,~0\,$m/s$) to test pure AM rendering ($PureAM$), and maintained the same ratio of $\frac{v}{d}$ for the other three combinations to have the same drawing frequency at 191\,Hz.

We rendered the ultrasound Tactons using the STRATOS Explore device by Ultraleap and measured the Tactons using a 1D laser vibrometer (Ploytec IVS-500) on paper (Figure~\ref{fig:setup}).
Initially, we tried to measure the displacements induced by the ultrasound device on the skin of palm but the induced displacement was too weak and lower than the vibrometer's minimum resolution.
After much experimentation, we configured a system to measure displacements induced by the ultrasound Tactons on a thin paper.
For the Tactons varying on \textit{AM frequency}, \textit{size}, and \textit{drawing speed}, 
we measured vibrations at 5 points inside, on the trajectory, and outside the shape, spaced at 5 mm intervals (Figure~\ref{fig:Mpoint}).

\subsection{Results}
The preliminary measurement results showed similar results to those reported in~\cite{chilles2019laser}, although we measured the vibrations on paper.
$PureAM$ introduced frequency harmonics (multiples of AM frequency or $f_{AM}$) (Figure~\ref{fig:Comparison}).
$PureSTM$ rendering showed frequency harmonics (multiples of drawing frequency or $f_{d}$), which appeared consistently for different drawing speed and size as long as the drawing frequency (the ratio of speed to size) was the same.
The combinations of AM and STM ($AM{+}STM$) also resulted in frequency harmonics at multiples of both AM frequency and drawing frequency. 
Also, the highest magnitude occurred at the inputted AM frequency among all harmonic frequencies, regardless of the STM parameters.

Our model for $PureAM$ and $PureSTM$ simulated the exact AM frequency and drawing frequency ($f_{AM}$ and $f_{d}$) as observed in the measurements (Figure~\ref{fig:Comparison}).
In particular, simulations for $PureSTM$ showed the same frequency harmonics at multiples of the drawing frequency.
However, for both $PureAM$ and $AM{+}STM$, the simulations and measurements showed less correspondence, perhaps due to the different characteristics between paper and human skin.
The simulated waveform for $PureAM$ included a single component at the inputted $f_{AM}$ in the spectral domain, while the measured waveform for $PureAM$ introduced frequency harmonics at multiples of $f_{AM}$ in the measurement.
In $AM{+}STM$, both the simulations and measurements showed the highest magnitude at the inputted AM frequency, regardless of the STM parameters.
However, the simulations showed the $f_{AM}$ component and pair frequency components at multiples of $f_{d} \pm f_{AM}$, while the measurements included frequency harmonics at multiples of both $f_{AM}$ and $f_{d}$. 
These frequency harmonics have a much lower magnitude than the AM frequency, so the extent of their impact on user perception is not fully known.

\section{Discussion}
\label{sec:discussion}

Our simulation can provide insights for designing mid-air ultrasound Tactons, as the designer can visualize the complex frequency spectrum induced by $PureSTM$ or $AM{+}STM$ and anticipate its impact on the end-user experience of the Tactons.
In other words, our proposed simulation can inform actual stimulation at any points above the device, aiding as an effective means for testing the physical effects of the created Tactons by haptic designers.

The preliminary measurement data suggested the limitations of the current measurement setup and simulation model.
At the current setup, the paper had different characteristics from human skin, such as elasticity and shear wave propagation, thus this measurement did not perfectly reflect the stimulation process on the human skin.
Moreover, our model architecture was built on the simplified assumptions (Section~\ref{sec:assumptions}) and on data collected from a scale hung in the air~\cite{raza2019perceptually} and a microphone placed away from the device~\cite{carter2013ultrahaptics}, instead of using human skin stimulation data.

In the future, we aim to improve the measurement methodology, for example, by using a higher resolution laser vibrometer or employing a silicon-based replica of human skin~\cite{frier2018using, chilles2019laser}. 
Also, we plan to improve the simulation's accuracy by relaxing our current assumption such as the propagation of vibration waves on the skin to fully capture the complexity of mid-air ultrasound stimulation.
Finally, we plan to improve our interactive simulation tool by including more design parameters, such as rhythmic structure and the number of focal points, and by adding the borderline of perceptible intensities considering detection threshold.
We plan to make our simulation tool open-source after making the above improvements and further validating the model with a larger set of Tactons.

\section{Conclusion}

A physical simulation for skin vibrations can offer new possibilities for rapidly prototyping mid-air ultrasound Tactons for user applications.
We proposed an interactive simulation for mid-air ultrasound Tactons that allows designers to easily test combinations of eight ultrasound parameters (five temporal and three spatiotemporal) and visualize the vibrations induced at different points on the skin. 
Our initial results suggest high correspondence between our simulation and measurements of a set of Tactons.
We hope the simulation can assist haptic designers in creating rich mid-air ultrasound Tactons that vary on temporal and spatial parameters.


\newpage
\newpage

\begin{acks}
We would like to thank Kyuyoung Shim and Gyungmin Jin for assisting with the preliminary measurements.
This work was supported by research grants from VILLUM FONDEN (VIL50296), the National Science Foundation (\#2339707), the Institute of Information \& communications Technology Planning \& Evaluation (IITP) funded by the Korea government (MSIT) (No.2019-0-01842, Artificial Intelligence Graduate School Program (GIST)), and the Culture, Sports and Tourism R\&D Program through the Korea Creative Content Agency funded by the Ministry of Culture, Sports and Tourism in 2023 (RS-2023-00226263).
\end{acks}

\bibliographystyle{ACM-Reference-Format}
\bibliography{99_Bibliography}


\begin{thebibliography}{35}


\ifx \showCODEN    \undefined \def \showCODEN     #1{\unskip}     \fi
\ifx \showDOI      \undefined \def \showDOI       #1{#1}\fi
\ifx \showISBNx    \undefined \def \showISBNx     #1{\unskip}     \fi
\ifx \showISBNxiii \undefined \def \showISBNxiii  #1{\unskip}     \fi
\ifx \showISSN     \undefined \def \showISSN      #1{\unskip}     \fi
\ifx \showLCCN     \undefined \def \showLCCN      #1{\unskip}     \fi
\ifx \shownote     \undefined \def \shownote      #1{#1}          \fi
\ifx \showarticletitle \undefined \def \showarticletitle #1{#1}   \fi
\ifx \showURL      \undefined \def \showURL       {\relax}        \fi
\providecommand\bibfield[2]{#2}
\providecommand\bibinfo[2]{#2}
\providecommand\natexlab[1]{#1}
\providecommand\showeprint[2][]{arXiv:#2}

\bibitem[Brown et~al\mbox{.}(2022)]%
        {brown2022augmenting}
\bibfield{author}{\bibinfo{person}{Eddie Brown}, \bibinfo{person}{David~R Large}, \bibinfo{person}{Hannah Limerick}, \bibinfo{person}{William Frier}, {and} \bibinfo{person}{Gary Burnett}.} \bibinfo{year}{2022}\natexlab{}.
\newblock \showarticletitle{Augmenting automotive gesture infotainment interfaces through mid-air haptic icon design}.
\newblock In \bibinfo{booktitle}{\emph{Ultrasound Mid-Air Haptics for Touchless Interfaces}}. \bibinfo{publisher}{Springer}, \bibinfo{pages}{119--145}.
\newblock


\bibitem[Carter et~al\mbox{.}(2013)]%
        {carter2013ultrahaptics}
\bibfield{author}{\bibinfo{person}{Tom Carter}, \bibinfo{person}{Sue~Ann Seah}, \bibinfo{person}{Benjamin Long}, \bibinfo{person}{Bruce Drinkwater}, {and} \bibinfo{person}{Sriram Subramanian}.} \bibinfo{year}{2013}\natexlab{}.
\newblock \showarticletitle{UltraHaptics: multi-point mid-air haptic feedback for touch surfaces}. In \bibinfo{booktitle}{\emph{Proceedings of the 26th annual ACM symposium on User interface software and technology}}. \bibinfo{pages}{505--514}.
\newblock


\bibitem[Chilles et~al\mbox{.}(2019)]%
        {chilles2019laser}
\bibfield{author}{\bibinfo{person}{Jamie Chilles}, \bibinfo{person}{William Frier}, \bibinfo{person}{Abdenaceur Abdouni}, \bibinfo{person}{Marcello Giordano}, {and} \bibinfo{person}{Orestis Georgiou}.} \bibinfo{year}{2019}\natexlab{}.
\newblock \showarticletitle{Laser doppler vibrometry and fem simulations of ultrasonic mid-air haptics}. In \bibinfo{booktitle}{\emph{2019 IEEE World Haptics Conference (WHC)}}. IEEE, \bibinfo{pages}{259--264}.
\newblock


\bibitem[Dalsgaard et~al\mbox{.}(2022)]%
        {dalsgaard2022user}
\bibfield{author}{\bibinfo{person}{Tor-Salve Dalsgaard}, \bibinfo{person}{Joanna Bergstr{\"o}m}, \bibinfo{person}{Marianna Obrist}, {and} \bibinfo{person}{Kasper Hornb{\ae}k}.} \bibinfo{year}{2022}\natexlab{}.
\newblock \showarticletitle{A user-derived mapping for mid-air haptic experiences}.
\newblock \bibinfo{journal}{\emph{International Journal of Human-Computer Studies}}  \bibinfo{volume}{168} (\bibinfo{year}{2022}), \bibinfo{pages}{102920}.
\newblock


\bibitem[Freeman and Wilson(2021)]%
        {freeman2021perception}
\bibfield{author}{\bibinfo{person}{Euan Freeman} {and} \bibinfo{person}{Graham Wilson}.} \bibinfo{year}{2021}\natexlab{}.
\newblock \showarticletitle{Perception of ultrasound haptic focal point motion}. In \bibinfo{booktitle}{\emph{Proceedings of the 2021 International Conference on Multimodal Interaction}}. \bibinfo{pages}{697--701}.
\newblock


\bibitem[Frier et~al\mbox{.}(2018)]%
        {frier2018using}
\bibfield{author}{\bibinfo{person}{William Frier}, \bibinfo{person}{Damien Ablart}, \bibinfo{person}{Jamie Chilles}, \bibinfo{person}{Benjamin Long}, \bibinfo{person}{Marcello Giordano}, \bibinfo{person}{Marianna Obrist}, {and} \bibinfo{person}{Sriram Subramanian}.} \bibinfo{year}{2018}\natexlab{}.
\newblock \showarticletitle{Using spatiotemporal modulation to draw tactile patterns in mid-air}. In \bibinfo{booktitle}{\emph{Haptics: Science, Technology, and Applications: 11th International Conference, EuroHaptics 2018, Pisa, Italy, June 13-16, 2018, Proceedings, Part I 11}}. Springer, \bibinfo{pages}{270--281}.
\newblock


\bibitem[Hajas et~al\mbox{.}(2020)]%
        {hajas2020mid}
\bibfield{author}{\bibinfo{person}{Daniel Hajas}, \bibinfo{person}{Dario Pittera}, \bibinfo{person}{Antony Nasce}, \bibinfo{person}{Orestis Georgiou}, {and} \bibinfo{person}{Marianna Obrist}.} \bibinfo{year}{2020}\natexlab{}.
\newblock \showarticletitle{Mid-air haptic rendering of 2D geometric shapes with a dynamic tactile pointer}.
\newblock \bibinfo{journal}{\emph{IEEE transactions on haptics}} \bibinfo{volume}{13}, \bibinfo{number}{4} (\bibinfo{year}{2020}), \bibinfo{pages}{806--817}.
\newblock


\bibitem[Harrington et~al\mbox{.}(2018)]%
        {harrington2018exploring}
\bibfield{author}{\bibinfo{person}{Kyle Harrington}, \bibinfo{person}{David~R Large}, \bibinfo{person}{Gary Burnett}, {and} \bibinfo{person}{Orestis Georgiou}.} \bibinfo{year}{2018}\natexlab{}.
\newblock \showarticletitle{Exploring the use of mid-air ultrasonic feedback to enhance automotive user interfaces}. In \bibinfo{booktitle}{\emph{Proceedings of the 10th international conference on automotive user interfaces and interactive vehicular applications}}. \bibinfo{pages}{11--20}.
\newblock


\bibitem[Hoshi et~al\mbox{.}(2009)]%
        {hoshi2009non}
\bibfield{author}{\bibinfo{person}{Takayuki Hoshi}, \bibinfo{person}{Takayuki Iwamoto}, {and} \bibinfo{person}{Hiroyuki Shinoda}.} \bibinfo{year}{2009}\natexlab{}.
\newblock \showarticletitle{Non-contact tactile sensation synthesized by ultrasound transducers}. In \bibinfo{booktitle}{\emph{World Haptics 2009-Third Joint EuroHaptics conference and Symposium on Haptic Interfaces for Virtual Environment and Teleoperator Systems}}. IEEE, \bibinfo{pages}{256--260}.
\newblock


\bibitem[Howard et~al\mbox{.}(2023)]%
        {howard2023gap}
\bibfield{author}{\bibinfo{person}{Thomas Howard}, \bibinfo{person}{Karina Driller}, \bibinfo{person}{William Frier}, \bibinfo{person}{Claudio Pacchierotti}, \bibinfo{person}{Maud Marchal}, {and} \bibinfo{person}{Jessica Hartcher-O’Brien}.} \bibinfo{year}{2023}\natexlab{}.
\newblock \showarticletitle{Gap detection in pairs of ultrasound mid-air vibrotactile stimuli}.
\newblock \bibinfo{journal}{\emph{ACM Transactions on Applied Perceptions}} \bibinfo{volume}{20}, \bibinfo{number}{1} (\bibinfo{year}{2023}), \bibinfo{pages}{1--17}.
\newblock


\bibitem[Howard et~al\mbox{.}(2019)]%
        {howard2019investigating}
\bibfield{author}{\bibinfo{person}{Thomas Howard}, \bibinfo{person}{Gerard Gallagher}, \bibinfo{person}{Anatole L{\'e}cuyer}, \bibinfo{person}{Claudio Pacchierotti}, {and} \bibinfo{person}{Maud Marchal}.} \bibinfo{year}{2019}\natexlab{}.
\newblock \showarticletitle{Investigating the recognition of local shapes using mid-air ultrasound haptics}. In \bibinfo{booktitle}{\emph{2019 IEEE World Haptics Conference (WHC)}}. IEEE, \bibinfo{pages}{503--508}.
\newblock


\bibitem[Howard et~al\mbox{.}(2022)]%
        {howard2022ultrasound}
\bibfield{author}{\bibinfo{person}{Thomas Howard}, \bibinfo{person}{Maud Marchal}, {and} \bibinfo{person}{Claudio Pacchierotti}.} \bibinfo{year}{2022}\natexlab{}.
\newblock \showarticletitle{Ultrasound mid-air tactile feedback for immersive virtual reality interaction}.
\newblock In \bibinfo{booktitle}{\emph{Ultrasound Mid-Air Haptics for Touchless Interfaces}}. \bibinfo{publisher}{Springer}, \bibinfo{pages}{147--183}.
\newblock


\bibitem[Hung et~al\mbox{.}(2013)]%
        {hung2013ultrapulse}
\bibfield{author}{\bibinfo{person}{Gary~MY Hung}, \bibinfo{person}{Nigel~W John}, \bibinfo{person}{Chris Hancock}, \bibinfo{person}{Derek~A Gould}, {and} \bibinfo{person}{Takayuki Hoshi}.} \bibinfo{year}{2013}\natexlab{}.
\newblock \showarticletitle{UltraPulse-simulating a human arterial pulse with focussed airborne ultrasound}. In \bibinfo{booktitle}{\emph{2013 35th Annual International Conference of the IEEE Engineering in Medicine and Biology Society (EMBC)}}. IEEE, \bibinfo{pages}{2511--2514}.
\newblock


\bibitem[Hung et~al\mbox{.}(2014)]%
        {hung2014using}
\bibfield{author}{\bibinfo{person}{Gary~MY Hung}, \bibinfo{person}{Nigel~W John}, \bibinfo{person}{Chris Hancock}, {and} \bibinfo{person}{Takayuki Hoshi}.} \bibinfo{year}{2014}\natexlab{}.
\newblock \showarticletitle{Using and validating airborne ultrasound as a tactile interface within medical training simulators}. In \bibinfo{booktitle}{\emph{Biomedical Simulation: 6th International Symposium, ISBMS 2014, Strasbourg, France, October 16-17, 2014. Proceedings 6}}. Springer, \bibinfo{pages}{30--39}.
\newblock


\bibitem[Hwang et~al\mbox{.}(2017a)]%
        {hwang2017perceptual}
\bibfield{author}{\bibinfo{person}{Inwook Hwang}, \bibinfo{person}{Jeongil Seo}, {and} \bibinfo{person}{Seungmoon Choi}.} \bibinfo{year}{2017}\natexlab{a}.
\newblock \showarticletitle{Perceptual space of superimposed dual-frequency vibrations in the hands}.
\newblock \bibinfo{journal}{\emph{PloS one}} \bibinfo{volume}{12}, \bibinfo{number}{1} (\bibinfo{year}{2017}), \bibinfo{pages}{e0169570}.
\newblock


\bibitem[Hwang et~al\mbox{.}(2017b)]%
        {hwang2017airpiano}
\bibfield{author}{\bibinfo{person}{Inwook Hwang}, \bibinfo{person}{Hyungki Son}, {and} \bibinfo{person}{Jin~Ryong Kim}.} \bibinfo{year}{2017}\natexlab{b}.
\newblock \showarticletitle{AirPiano: Enhancing music playing experience in virtual reality with mid-air haptic feedback}. In \bibinfo{booktitle}{\emph{2017 IEEE world haptics conference (WHC)}}. IEEE, \bibinfo{pages}{213--218}.
\newblock


\bibitem[Lim and Park(2023)]%
        {lim2023can}
\bibfield{author}{\bibinfo{person}{Chungman Lim} {and} \bibinfo{person}{Gunhyuk Park}.} \bibinfo{year}{2023}\natexlab{}.
\newblock \showarticletitle{Can a Computer Tell Differences between Vibrations?: Physiology-Based Computational Model for Perceptual Dissimilarity Prediction}. In \bibinfo{booktitle}{\emph{Proceedings of the 2023 CHI Conference on Human Factors in Computing Systems}}. \bibinfo{pages}{1--15}.
\newblock


\bibitem[Limerick et~al\mbox{.}(2019)]%
        {limerick2019user}
\bibfield{author}{\bibinfo{person}{Hannah Limerick}, \bibinfo{person}{Richard Hayden}, \bibinfo{person}{David Beattie}, \bibinfo{person}{Orestis Georgiou}, {and} \bibinfo{person}{J{\"o}rg M{\"u}ller}.} \bibinfo{year}{2019}\natexlab{}.
\newblock \showarticletitle{User engagement for mid-air haptic interactions with digital signage}. In \bibinfo{booktitle}{\emph{Proceedings of the 8th ACM international symposium on pervasive displays}}. \bibinfo{pages}{1--7}.
\newblock


\bibitem[Mulot et~al\mbox{.}(2023a)]%
        {mulot2023improving}
\bibfield{author}{\bibinfo{person}{Lendy Mulot}, \bibinfo{person}{Thomas Howard}, \bibinfo{person}{Claudio Pacchierotti}, {and} \bibinfo{person}{Maud Marchal}.} \bibinfo{year}{2023}\natexlab{a}.
\newblock \showarticletitle{Improving the Perception of Mid-Air Tactile Shapes With Spatio-Temporally-Modulated Tactile Pointers}.
\newblock \bibinfo{journal}{\emph{ACM Transactions on Applied Perception}} \bibinfo{volume}{20}, \bibinfo{number}{4} (\bibinfo{year}{2023}), \bibinfo{pages}{1--16}.
\newblock


\bibitem[Mulot et~al\mbox{.}(2023b)]%
        {mulot2023ultrasound}
\bibfield{author}{\bibinfo{person}{Lendy Mulot}, \bibinfo{person}{Thomas Howard}, \bibinfo{person}{Claudio Pacchierotti}, {and} \bibinfo{person}{Maud Marchal}.} \bibinfo{year}{2023}\natexlab{b}.
\newblock \showarticletitle{Ultrasound Mid-Air Haptics for Hand Guidance in Virtual Reality}.
\newblock \bibinfo{journal}{\emph{IEEE Transactions on Haptics}} (\bibinfo{year}{2023}).
\newblock


\bibitem[Obrist et~al\mbox{.}(2013)]%
        {obrist2013talking}
\bibfield{author}{\bibinfo{person}{Marianna Obrist}, \bibinfo{person}{Sue~Ann Seah}, {and} \bibinfo{person}{Sriram Subramanian}.} \bibinfo{year}{2013}\natexlab{}.
\newblock \showarticletitle{Talking about tactile experiences}. In \bibinfo{booktitle}{\emph{Proceedings of the SIGCHI conference on human factors in computing systems}}. \bibinfo{pages}{1659--1668}.
\newblock


\bibitem[Obrist et~al\mbox{.}(2015)]%
        {obrist2015emotions}
\bibfield{author}{\bibinfo{person}{Marianna Obrist}, \bibinfo{person}{Sriram Subramanian}, \bibinfo{person}{Elia Gatti}, \bibinfo{person}{Benjamin Long}, {and} \bibinfo{person}{Thomas Carter}.} \bibinfo{year}{2015}\natexlab{}.
\newblock \showarticletitle{Emotions mediated through mid-air haptics}. In \bibinfo{booktitle}{\emph{Proceedings of the 33rd annual ACM conference on human factors in computing systems}}. \bibinfo{pages}{2053--2062}.
\newblock


\bibitem[Park and Choi(2011)]%
        {park2011perceptual}
\bibfield{author}{\bibinfo{person}{Gunhyuk Park} {and} \bibinfo{person}{Seungmoon Choi}.} \bibinfo{year}{2011}\natexlab{}.
\newblock \showarticletitle{Perceptual space of amplitude-modulated vibrotactile stimuli}. In \bibinfo{booktitle}{\emph{2011 IEEE world haptics conference}}. IEEE, \bibinfo{pages}{59--64}.
\newblock


\bibitem[Rakkolainen et~al\mbox{.}(2020)]%
        {rakkolainen2020survey}
\bibfield{author}{\bibinfo{person}{Ismo Rakkolainen}, \bibinfo{person}{Euan Freeman}, \bibinfo{person}{Antti Sand}, \bibinfo{person}{Roope Raisamo}, {and} \bibinfo{person}{Stephen Brewster}.} \bibinfo{year}{2020}\natexlab{}.
\newblock \showarticletitle{A survey of mid-air ultrasound haptics and its applications}.
\newblock \bibinfo{journal}{\emph{IEEE Transactions on Haptics}} \bibinfo{volume}{14}, \bibinfo{number}{1} (\bibinfo{year}{2020}), \bibinfo{pages}{2--19}.
\newblock


\bibitem[Raza et~al\mbox{.}(2019)]%
        {raza2019perceptually}
\bibfield{author}{\bibinfo{person}{Ahsan Raza}, \bibinfo{person}{Waseem Hassan}, \bibinfo{person}{Tatyana Ogay}, \bibinfo{person}{Inwook Hwang}, {and} \bibinfo{person}{Seokhee Jeon}.} \bibinfo{year}{2019}\natexlab{}.
\newblock \showarticletitle{Perceptually correct haptic rendering in mid-air using ultrasound phased array}.
\newblock \bibinfo{journal}{\emph{IEEE Transactions on Industrial Electronics}} \bibinfo{volume}{67}, \bibinfo{number}{1} (\bibinfo{year}{2019}), \bibinfo{pages}{736--745}.
\newblock


\bibitem[Rutten et~al\mbox{.}(2020)]%
        {rutten2020discriminating}
\bibfield{author}{\bibinfo{person}{Isa Rutten}, \bibinfo{person}{William Frier}, {and} \bibinfo{person}{David Geerts}.} \bibinfo{year}{2020}\natexlab{}.
\newblock \showarticletitle{Discriminating between intensities and velocities of mid-air haptic patterns}. In \bibinfo{booktitle}{\emph{Haptics: Science, Technology, Applications: 12th International Conference, EuroHaptics 2020, Leiden, The Netherlands, September 6--9, 2020, Proceedings 12}}. Springer, \bibinfo{pages}{78--86}.
\newblock


\bibitem[Rutten et~al\mbox{.}(2019)]%
        {rutten2019invisible}
\bibfield{author}{\bibinfo{person}{Isa Rutten}, \bibinfo{person}{William Frier}, \bibinfo{person}{Lawrence Van~den Bogaert}, {and} \bibinfo{person}{David Geerts}.} \bibinfo{year}{2019}\natexlab{}.
\newblock \showarticletitle{Invisible touch: How identifiable are mid-air haptic shapes?}. In \bibinfo{booktitle}{\emph{Extended abstracts of the 2019 CHI conference on human factors in computing systems}}. \bibinfo{pages}{1--6}.
\newblock


\bibitem[Seifi et~al\mbox{.}(2015)]%
        {seifi2015vibviz}
\bibfield{author}{\bibinfo{person}{Hasti Seifi}, \bibinfo{person}{Kailun Zhang}, {and} \bibinfo{person}{Karon~E MacLean}.} \bibinfo{year}{2015}\natexlab{}.
\newblock \showarticletitle{VibViz: Organizing, visualizing and navigating vibration libraries}. In \bibinfo{booktitle}{\emph{2015 IEEE World Haptics Conference (WHC)}}. IEEE, \bibinfo{pages}{254--259}.
\newblock


\bibitem[Shen et~al\mbox{.}(2023)]%
        {shen2023multi}
\bibfield{author}{\bibinfo{person}{Zhouyang Shen}, \bibinfo{person}{Madhan~Kumar Vasudevan}, \bibinfo{person}{Jan Ku{\v{c}}era}, \bibinfo{person}{Marianna Obrist}, {and} \bibinfo{person}{Diego Martinez~Plasencia}.} \bibinfo{year}{2023}\natexlab{}.
\newblock \showarticletitle{Multi-point STM: Effects of Drawing Speed and Number of Focal Points on Users’ Responses using Ultrasonic Mid-Air Haptics}. In \bibinfo{booktitle}{\emph{Proceedings of the 2023 CHI Conference on Human Factors in Computing Systems}}. \bibinfo{pages}{1--11}.
\newblock


\bibitem[Takahashi et~al\mbox{.}(2019)]%
        {takahashi2019tactile}
\bibfield{author}{\bibinfo{person}{Ryoko Takahashi}, \bibinfo{person}{Keisuke Hasegawa}, {and} \bibinfo{person}{Hiroyuki Shinoda}.} \bibinfo{year}{2019}\natexlab{}.
\newblock \showarticletitle{Tactile stimulation by repetitive lateral movement of midair ultrasound focus}.
\newblock \bibinfo{journal}{\emph{IEEE transactions on haptics}} \bibinfo{volume}{13}, \bibinfo{number}{2} (\bibinfo{year}{2019}), \bibinfo{pages}{334--342}.
\newblock


\bibitem[Vi et~al\mbox{.}(2017)]%
        {vi2017not}
\bibfield{author}{\bibinfo{person}{Chi~Thanh Vi}, \bibinfo{person}{Damien Ablart}, \bibinfo{person}{Elia Gatti}, \bibinfo{person}{Carlos Velasco}, {and} \bibinfo{person}{Marianna Obrist}.} \bibinfo{year}{2017}\natexlab{}.
\newblock \showarticletitle{Not just seeing, but also feeling art: Mid-air haptic experiences integrated in a multisensory art exhibition}.
\newblock \bibinfo{journal}{\emph{International Journal of Human-Computer Studies}}  \bibinfo{volume}{108} (\bibinfo{year}{2017}), \bibinfo{pages}{1--14}.
\newblock


\bibitem[Villa et~al\mbox{.}(2022)]%
        {villa2022extended}
\bibfield{author}{\bibinfo{person}{Steeven Villa}, \bibinfo{person}{Sven Mayer}, \bibinfo{person}{Jess Hartcher-O'Brien}, \bibinfo{person}{Albrecht Schmidt}, {and} \bibinfo{person}{Tonja-Katrin Machulla}.} \bibinfo{year}{2022}\natexlab{}.
\newblock \showarticletitle{Extended mid-air ultrasound haptics for virtual reality}.
\newblock \bibinfo{journal}{\emph{Proceedings of the ACM on Human-Computer Interaction}} \bibinfo{volume}{6}, \bibinfo{number}{ISS} (\bibinfo{year}{2022}), \bibinfo{pages}{500--524}.
\newblock


\bibitem[Wilson et~al\mbox{.}(2014)]%
        {wilson2014perception}
\bibfield{author}{\bibinfo{person}{Graham Wilson}, \bibinfo{person}{Thomas Carter}, \bibinfo{person}{Sriram Subramanian}, {and} \bibinfo{person}{Stephen~A Brewster}.} \bibinfo{year}{2014}\natexlab{}.
\newblock \showarticletitle{Perception of ultrasonic haptic feedback on the hand: localisation and apparent motion}. In \bibinfo{booktitle}{\emph{Proceedings of the SIGCHI conference on human factors in computing systems}}. \bibinfo{pages}{1133--1142}.
\newblock


\bibitem[Wojna et~al\mbox{.}(2023)]%
        {wojna2023exploration}
\bibfield{author}{\bibinfo{person}{Katarzyna Wojna}, \bibinfo{person}{Orestis Georgiou}, \bibinfo{person}{David Beattie}, \bibinfo{person}{William Frier}, \bibinfo{person}{Michael Wright}, {and} \bibinfo{person}{Christof Lutteroth}.} \bibinfo{year}{2023}\natexlab{}.
\newblock \showarticletitle{An Exploration of Just Noticeable Differences in Mid-Air Haptics}. In \bibinfo{booktitle}{\emph{2023 IEEE World Haptics Conference (WHC)}}. IEEE, \bibinfo{pages}{410--416}.
\newblock


\bibitem[Yoo et~al\mbox{.}(2022)]%
        {yoo2022perceived}
\bibfield{author}{\bibinfo{person}{Yongjae Yoo}, \bibinfo{person}{Inwook Hwang}, {and} \bibinfo{person}{Seungmoon Choi}.} \bibinfo{year}{2022}\natexlab{}.
\newblock \showarticletitle{Perceived intensity model of dual-frequency superimposed vibration: Pythagorean sum}.
\newblock \bibinfo{journal}{\emph{IEEE Transactions on Haptics}} \bibinfo{volume}{15}, \bibinfo{number}{2} (\bibinfo{year}{2022}), \bibinfo{pages}{405--415}.
\newblock


\end{thebibliography}











\end{document}